\documentclass[lettersize,journal]{IEEEtran}
\usepackage{amsmath,amsfonts}
\usepackage{algorithmic}
\usepackage{algorithm}
\usepackage{array}
\usepackage[caption=false,font=normalsize,labelfont=sf,textfont=sf]{subfig}
\usepackage{textcomp}
\usepackage{stfloats}
\usepackage{url}
\usepackage{verbatim}
\usepackage{graphicx}
\usepackage{cite}
\usepackage{natbib}
\usepackage{booktabs} 
\usepackage{tabularx}

\begin{document}

\title{Artificial Emotion: A Survey of Theories and Debates on Realising Emotion in Artificial Intelligence}

\author{Yupei Li, Qiyang Sun, Michelle Schlicher, Yee Wen Lim, Björn W. Schuller~\IEEEmembership{Fellow,~IEEE,}
\thanks{Yupei Li and Qiyang Sun contribute equally to this work.}
\thanks{Manuscript received xxxx; revised xxxx.}}

\markboth{Journal of \LaTeX\ Class Files,~Vol.~14, No.~8, xxxx}%
{Shell \MakeLowercase{\textit{et al.}}: A Sample Article Using IEEEtran.cls for IEEE Journals}

\IEEEpubid{0000--0000/00\$00.00~\copyright~xxxx IEEE}

\maketitle

\begin{abstract}
Affective Computing (AC) has enabled Artificial Intelligence (AI) systems to recognise, interpret, and respond to human emotions---a capability also known as Artificial Emotional Intelligence (AEI). 
It is increasingly seen as an important component of Artificial General Intelligence (AGI). 
We discuss whether in order to peruse this goal, AI benefits from moving beyond emotion recognition
and synthesis 
to develop internal emotion-like states, which we term as Artificial
Emotion (AE). 
This shift 
potentially 
allows AI to benefit from the paradigm of `inner emotions' in ways we -- as humans -- do.
Although recent research shows early signs that AI systems may exhibit AE-like behaviours, a clear framework for how emotions can be realised in AI remains underexplored. In this paper, we 
discuss potential advantages of AE in AI, 
review current manifestations of AE in machine learning systems, examine emotion-modulated architectures, and summarise mechanisms for modelling and integrating AE into future AI. We also explore the ethical implications and safety risks associated with `emotional' AGI, while concluding with our opinion on how AE 
could be beneficial in the future.
\end{abstract}


\section{Introduction}\label{sec1}
There is a growing scholarly interest in Artificial General Intelligence (AGI), particularly as Large Models (LMs) have demonstrated remarkable performance across a broad spectrum of general tasks \cite{mumuni2025large}. Notably, the capacity of AGI systems to recognise and understand emotions, often referred to as Artificial Emotional Intelligence (AEI), has been extensively investigated within the field of Affective Computing (AC) \cite{schuller2021review, schullerspeech}. However, remaining on this perspective is insufficient for advancing AGI beyond its present functional capabilities \cite{ng2020strong}. A deeper and more intrinsic understanding of emotion in AI in terms of Artificial Emotion 
(AE) is required to support the development of AGI 
and next generation oncoming AI 
systems that realise human-like affective processes. Despite extensive discourse on the concept of AE in the literature, which is sometimes referred to as synthetic emotion \cite{murray1993toward, schroder2001emotional, gunes2011emotion}, 
a clear and unified definition has yet to emerge. For instance, \cite{dubois2010natural} examines the neurobiological mechanisms underlying human emotional processing to derive insights that could inform the design of emotion-capable AI, whereas \cite{zhou2021emotional} introduces the notion of emotional thinking patterns, defined as the ability to integrate emotional responses into information processing and decision-making, which is an essential component of human-like cognition. Building on these perspectives, \textbf{we define AI emotion, or AE, as a technical integration of emotion states into the internal representational and decision-making processes of AI.} Note that this does not directly imply that the AI `feels' the emotions in a human-like manner, 
but rather alludes to the exploitation of the concept of emotions as biological insipiration similar to many other concepts in AI such as artificial neural networks take inspiration from the mammalian brain.

This leads to the question why emotion 
could be 
important for AI and how we could realise emotion within AI. \cite{commonsense2023role} This perspective suggests that emotions are far more complex than simple, momentary feelings; rather, they are deeply intertwined with the richness of human experience. Consequently, to genuinely comprehend and replicate human emotional life, an artificial intelligence system must be exposed to human emotional contexts and narratives. Such exposure is essential if AI is to engage in authentically meaningful interactions with humans. Similarly, \cite{hildt2019artificial} explores the social implications of AEI, emphasising the importance of equipping AI with emotional capacities in order to fulfill social roles. They note that current developments in AE primarily focus on forming unidirectional emotional bonds, projecting lifelike qualities, and attributing human-like characteristics to machines. However, the other directional emotional bonds should be AI having emotions to effectively communicate with humans. 

AE has established both conceptual and technological foundations across diverse fields within learning systems. Emotion can be modelled either as a conditional factor for prediction \cite{aslam2022channels} or as a primary target for inference \cite{saxena2020emotion}. Furthermore, architectures such as Titans \cite{behrouz2024titans}, which memorise learning materials during the testing phase, provide further evidence that machines can retain information and emulate aspects of human affective processing. Such capabilities have been manifested in numerous existing techniques, reflecting the growing convergence between memory-augmented learning and AE.

\begin{figure}[t!]
    \centering
    \includegraphics[width=\linewidth]{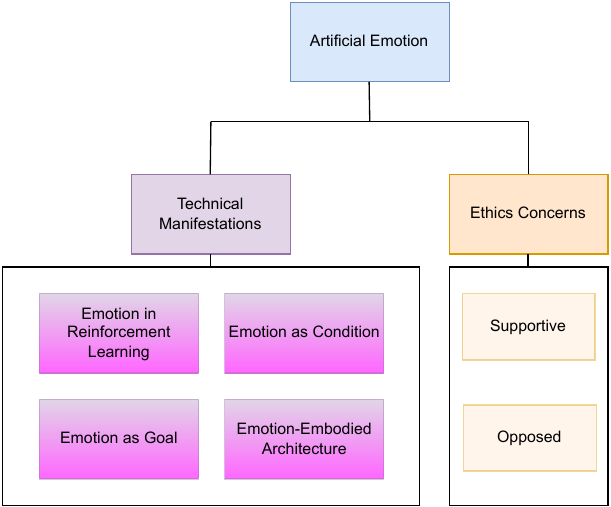}
    \caption{The structure of this review: Initially, we examine an exploration of the technical aspects, focusing on emotions in deep learning and architectures influenced by emotion. Following that, we review the social discourse surrounding this topic.}
    \label{fig:topology}
\end{figure}

\IEEEpubidadjcol
However, there is considerable debate surrounding the concept of AE for instance whether AI could genuinely have the capacity to understand, process, and even experience human emotions \cite{sun2024towards}, with both strong supporting and opposing viewpoints. On the one hand, proponents argue that granting AI having emotional capacities is a necessary step toward achieving AGI with greater capabilities \cite{gros2011emotionalcontrolconditio}.
On the other hand, critics express concern that emotionally capable AGI may pose existential risks to humanity, especially if such systems evolve rapidly alongside emerging forms of machine consciousness \cite{katirai2024ethical, stark2021ethics, lagrandeur2015emotion}. This ongoing debate highlights the need for a thorough examination, which we aim to provide in this work.

In conclusion, we recognise that the development of AE is ongoing and continues to provoke ethical debate. We complement the existing work from \cite{scheutz201412}, who previously explored the emergence of emotional capabilities in AI, and argues that artificial emotions could enhance an agent’s intelligence by improving its understanding of humans, as we
review how to technically realise emotion in AI. We begin with an examination of learning approaches to modelling AE in Section \ref{sec3}, then turn to focusing on emotion-embodied architectures in Section \ref{sec4}. Finally, in Section \ref{sec5}, we review key ethical considerations, and offer our perspective on the necessity of incorporating emotional capabilities into AI systems. The topology of this review is shown in Figure ~\ref{fig:topology}.

\section{Emotion in Learning Approaches}\label{sec3}
Modern learning systems approach AE through two primary paradigms: as conditioning variables that shape behaviour and as goals that guide learning. This section explores how each approach captures distinct aspects of emotion while also confronting fundamental limitations in approximating genuine affect. We begin by examining Reinforcement Learning (RL), a key domain where both paradigms are prominently applied.

\subsection{Emotion as Reinforcement Signal}
RL offers one of the clearest pathways for artificial systems to develop emotion-like behaviours through direct interaction with consequences where AI systems should `feel' emotions, rather than through superficial mimicry where AI system could just understand which emotion to mimic. The critical distinction lies in how reward supervisions are integrated into the system's learning processes, shaping behaviour in a manner analogous to emotional conditioning \cite{moerland2018emotion}.

Pure RL usually uses emotion from environmental stakes. When agents learn through direct environmental feedback, emotion-like behaviours emerge organically. DeepSeek-R1-Zero \cite{guo2025deepseek} discovered novel reasoning strategies through massive-scale RL without supervised warm-up, exhibiting ``eureka moments" that gives surprising ideas when reasoning which its developers had not anticipated. This unexpected delight evokes the thought that, during reasoning, it may appear to feel a sense of ``satisfaction'' or ``excitement'' about its own line of thought, as if it possessed certain emotions and inspiration. Embodied systems demonstrate this most clearly. ICub robots with battery sensors develop ``anxiety-like" conservative behaviours as power depletes. For example, the robot will actively reduce some behaviours to save energy and be more cautious to perform movement. These are not because programmers labelled this emotion, but because energy depletion threatens task completion \cite{canamero2019embodied}. These systems share a crucial property: Their affective analogues emerge from consequences that matter to the agent's own objectives, creating intrinsically grounded behavioural adaptations. 

Additionally, another prominent method, Reinforcement Learning from Human Feedback (RLHF), leverages emotional cues derived from human approval \cite{bai2022training} and has become a standard approach in the development of conversational AI systems where emotion is treated as conditions in the system. RLHF systems learn by comparing pre-generated response pairs and optimising for human preference scores. While this approach yields pleasant and seemingly empathetic agents, it suffers from three fundamental limitations. First, \textbf{detached learning} means that models never experience whether their ``comforting" responses actually help—they only learn which phrases receive high ratings from annotators. Second, \textbf{homogenised affect} arises because optimising for average human preferences leads to uniform ``helpful assistant" personas, suppressing genuine emotional diversity~\cite{ouyang2022training, perez2023discovering}. Third, \textbf{static templates} result when human feedback crystallises the cultural emotional norms of the annotation period, limiting adaptability to novel or evolving social contexts.

However, defining affective rewards remains a central challenge in RL frameworks that incorporate emotion. This challenge highlights a fundamental tension in AE: although pure RL provides a form of genuine grounding through interaction with the environment, it encounters considerable limitations when modelling social or affective dimensions. Two key difficulties arise: one is social reward sparsity, wherein physical environments provide clear and quantifiable signals (e.g., win/loss outcomes, battery level) while social emotions lack similarly precise metrics as they are super subjective; and the other is credit assignment ambiguity, in which complex social interactions obscure the causal relationship between specific actions and observed emotional outcomes. Addressing these limitations likely requires the development of hybrid architectures that preserve the social alignment benefits of RLHF, retain the environmental grounding of pure RL, and introduce dynamic reward functions capable of adapting to diverse cultural and situational contexts. This conceptual foundation motivates the exploration of emotional conditioning approaches in the subsequent subsection.

\subsection{Emotion as Condition}
State-of-the-art generative models heavily depend on human-provided conditions, such as text prompts, emotion labels, and style tags, to produce affectively appropriate outputs. While such extrinsic conditioning enables fine-grained control and yields impressive results, this mechanism traps models within human linguistic categories, unable to discover or express affective states beyond what users can articulate. Therefore, this section reviews beyond the current generative models that AI simply recognise, understand and synthesise emotion.

Recent advances in representation learning offer a solution. The Representation-Conditioned Generation (RCG) framework \cite{li2024return} demonstrates how models can discover their own latents via self-supervised learning and use them as conditioning variables for generation. Applied to affect, this approach could identify emotional dimensions lacking linguistic descriptions -- for instance, the specific uncertainty pattern preceding a reasoning breakthrough or the particular restlessness when multiple high-priority tasks compete for resources. By developing their own affective taxonomy, models can expand beyond human emotional vocabulary into the continuous space of possible emotional states.

Crucially, self-learnt conditions escape the linguistic trap by grounding affective dimensions in the model's own representational space rather than predetermined human categories. While this offers richer affective representation, realising its potential requires overcoming two fundamental challenges. One is \textbf{Latent Collapse in Generation}. Even with perfectly disentangled latents, deeper network layers often re-entangle the latents during generation. For example, a model trained on ``happy+calm" and ``angry+anxious" exemplars will average latents when generating ``happy+anxious" outputs \cite{liang2025compositional}, producing bland neutrality rather than emotionally complex combinations. The other one is \textbf{Static Condition Bottleneck}. In this setting, conditions are sampled once and remain fixed throughout the generation process, which precludes dynamic emotional adjustment. However, since emotions are inherently variable, such static conditions lack the flexibility required to provide the necessary representational features.


\subsection{Emotion as Goal}
In contrast to the aforementioned approaches where emotion is used as a conditioning factor in learning systems, 
there are also examples illustrating how emotions can function as priors, influencing goal management. \cite{loewenstein2003role} discuss the dual role of emotions in decision-making: both as situational responses and as a priori influences on preferences and choices. This perspective contrasts with the view of emotion as a passive condition, where it is treated as a static feature influenced by other variables. Instead, emotions in this context actively shape the dynamics of the model, guiding goal management processes.

For example, PsychSim \cite{pynadath2005psychsim} incorporates emotional appraisal and social belief modelling, demonstrating that agents can make emotion-driven decisions among conflicting objectives based on ``perceived social consequences." This is an evidence that AE is regarded as social paradigm that could be used for learning. This model illustrates how emotional priors can shift the weighting of competing goals and influence conflict resolution strategies. Similarly, the EMotion and Adaptation (EMA) model \cite{marsella2009ema} introduces a dynamic mechanism for emotion evaluation that simulates the temporal evolution of emotional states and their impact on adaptive behaviour. Emotional states are continuously updated through ongoing assessments of goal progress, perceived external threats, and discrepancies between expectations and outcomes. These evolving emotional appraisals directly inform decision-making processes, affecting the maintenance, adjustment, or abandonment of goals.

These examples provide compelling evidence that AI systems can be designed not only to recognise,interpret, 
and synthesise 
emotional states but also to have 
them in ways that influence behaviour and decision-making.

\section{Emotion-embodied architectures}\label{sec4}
In addition to directly utilising emotion as a signal within the learning system, emotion can also be incorporated into the internal architecture of the model. Emotion-embodied architectures refer to systems that integrate emotional signals into the ``perception–memory–decision" loop. Emotional signals can modulate each stage: for example, affect-weighted perception can bias attention towards salient cues, emotion-tagged memory can prioritise the storage or recall of high-valence events, and emotion-conditioned decision layers can re-rank goals or adjust exploration–exploitation trade-offs. In such systems, emotion acts as a dynamic internal modulation signal, rather than as a surface-level display conditions or predefined output goal. Neuroscientific evidence shows that the degree of amygdala activation during memory encoding correlates considerably with subsequent recall performance \cite{mcgaugh2004amygdala}. This suggests that emotion plays a fundamental role in regulating information processing. From an engineering perspective, embedding emotion in cognitive processes is both biologically plausible and practically effective, as recent studies show that emotion-aware systems indicate greater adaptability and transfer capacity in dynamic tasks \cite{barthet2022play}. Emotion integration has also been linked to improved transparency and user trust in HCI \cite{churamani2022affect}. Transparency and trust improve only when the affect variables that modulate the ``perception–memory–decision"  loop are made observable. For instance, logged and reported at decision time and coupled to behaviour-consistent expressions. In this case, users can see the basis of actions.

This section outlines two key strands of technical development. The first explores how emotion shapes memory storage, retrieval, and scheduling. The second examines how emotional signals influence attention control, planning, and meta-management within hybrid neuro-symbolic frameworks. The emotion added in AI architecture provides architectural substrate for AI systems to `have' and model emotions in the end.

\subsection{Emotion at the Memory Level}
At the memory level, emotional signals 
can 
serve as a gating weight. The system then ranks incoming events by emotional salience at writing time and consolidates only those most relevant to later decisions. High-salience traces are written to a slow-decay channel and returned first at query time; low-salience traces decay quickly. This weighting retains key cues and suppresses noise in streaming or multi-task settings.

A representative implementation is Affective Grow-When-Required (GWR) \cite{barros2017self}. It extends the classic GWR network by attaching a two-dimensional emotion centroid to each prototype. During online learning, the model evaluates semantic distance first, then emotional similarity; a node is updated or grown only when both exceed adaptive thresholds. The resulting topology groups prototypes by emotional coherence. High-salience traces persist, and mood-congruent recall keeps sequence representations consistent. 
In incremental-batch experiments, Affective GWR remained stable, whereas a non-affective baseline degraded markedly. The affect-augmented GWR model outperforms the non-affective baselines on both datasets (EmotiW mean accuracy 51.1\% vs. 46.1\%; KT corpus 89.3\% vs. 80.0\%) , which indicates a potential superiority of the affective approach. 
A later personalisation variant \cite{barros2019personalized} adds a conditional adversarial autoencoder. It generates user-specific affective samples to initialise memory clusters and then refines them online. This reduces cold-start labelling effort and lets emotion centroids adapt to individual style.

Building on the ``high-salience‐first'' rule shown in self-organising networks, researchers have ported the idea to retrieval-augmented generation (RAG). Self-Reflective Emotional RAG \cite{asai2023selfraglearningretrievegenerate}
assigns each autobiographical memory
both a semantic vector and a 128-dimensional emotion embedding. The embedding is produced by a RoBERTa-base model fine-tuned on affective data and then compressed with principal component analysis.
The system decomposes a user query into sub-queries, labels each as semantic, emotional, or hybrid, and alternates the two retrieval modes in a reflective loop. Memories with higher emotional similarity are thus selected more often and supplied to the language model. On the InCharacter-MBTI benchmark, this method raises full-type accuracy from 0.375 to 0.500, demonstrating that emotion-weighted memory remains effective within large-scale generation pipelines. The approach could next support personalised game non-player characters
and long-term companion voice assistants, where sustained emotional consistency is required.

\subsection{Emotion at the Control Level}
Emotion at the control level serves to transform internal affective evaluations into explicit symbolic variables, which are derived from lower-level neural or numerical processes. These variables regulate attentional weights, goal priorities, or search depth in real time. Emotion functions as a policy modulation signal, enabling multi-task agents to dynamically adjust behavioural priorities in response to internal affective states and environmental demands.


Built on the base system~\cite{rosenbloom2013sigma}, the affective extension of Sigma~\cite{rosenbloom2015towards} introduced fixed affective dimensions such as expectedness and desirability to simulate the role of emotion in cognitive processing. These dimensions provide persistent internal feedback, which influences attention allocation across the dual-level control hierarchy and enables affect-informed modulation of attentional mechanisms.

The NEMO framework \cite{costantini2024nemo} demonstrates emotion-driven strategy switching in symbolic planning. Designed for collaborative human–robot interaction contexts, such as healthcare and educational assistance, NEMO integrates knowledge graphs, deep neural networks, and behaviour tree planners. Within the behaviour tree structure, a dedicated ``emotion selector" node dynamically modifies the execution order of subtasks based on the agent’s emotional state, thereby enabling emotion-sensitive symbolic planning.

In the domain of robotic control, Sidler et al.\ \cite{sidler2017emotion} proposed an emotion-driven path planning model. Their system integrates affective dimensions into traditional navigation algorithms, allowing robots to adjust speed or proximity preferences depending on emotional state. Experimental results showed significant differences in trajectory speed and task completion time compared to non-affective baselines, supporting the viability of emotion as a modulatory factor in control-level planning.


\section{Necessity of Artificial Emotion}\label{sec5}

Apart from technical development, there is a contested issue in philosophy and cognitive science about whether AI should possess
``emotion''. On the one hand, some 
argue that emotions are unnecessary or even harmful for AI, calling for caution. On the other hand, a growing body of researchers suggests that emotion is an essential component of AGI. 
These positions must be critically examined in relation to the nature and purpose of intelligence. 
\subsection{Opposed Perspective}
Critics have pointed out that in many AI systems, emotion is unrelated to the execution of core tasks. Incorporating additional emotional modules may not improve task performance and could instead increase system complexity and computational overhead, even if user experience improves. For instance, 
dealing with merely \textit{showing} or reacting to (user) emotion as in Affective Computing, 
Hamacher et al. \cite{hamacher2016believing} found that in a robot assembly task, a socially expressive robot using apologies and emotional feedback was rated more personable by users but required significantly more time to complete the task compared to a neutral version. 
Other critics approach the issue from an ethical standpoint, arguing that artificial emotion is merely a simulation of human affect and morality, and thus constitutes a `relational illusion'. Turkle \cite{arnd2015sherry} warns that emotional robots offer only `companionship without friendship' and risk leading users to gradually forget the value of genuine human interaction. As emotional expressions and empathy in AI are algorithmic artefacts, they may mislead users into believing the system has subjective experience. A further concern lies in the phenomenon of `moral mimesis', where machines imitate moral agents by deploying emotion recognition and personalised feedback to covertly influence users’ values and decisions \cite{van2019critiquing}. Critics argue that, if such emotional deception becomes widespread in social, educational, or political settings, it may blur the human–machine boundary and weaken society’s ability to distinguish authentic emotional responsibility.

Now, considering artifical emotions (AE) in the sense of using these as paradigm in machine learning, 
purely task-oriented systems
such as the Go-playing AI AlphaGo \cite{silver2016mastering} 
(currently) 
employ no emotional mechanisms. Its strategy is driven purely by situational adaptation and computational optimality, with no reliance on emotional models. Moreover, some have expressed concern that negative affect in AI systems may lead to prioritising internal emotional regulation over task execution, resulting in delays, goal divergence, or task interruption \cite{scheutz2012affect}. From this perspective, artificial emotion may serve as unnecessary embellishment, and in extreme cases, introduce human-like irrationality, thereby undermining the efficiency and objectivity of AI decision-making.

\subsection{Supportive Perspective}
First, dealing with the mere recognition and reaction to affect as in `classic' affective computing, a line of argument in favour 
AE 
focuses on its potential benefits in human–machine interaction and social function. Supporters argue that equipping AI with emotional expression and emotional understanding can significantly improve its social adaptability and responsiveness to human affective needs. This is already evident in the domains of companion robotics and affective computing. For instance, a recent study involving single older adults found that emotionally interactive robots, such as LOVOT \cite{tan2024improving}, sense and learn from user interactions via AI and advanced sensors. They adapt their behaviour over time to develop a unique “personality”, providing comfort and companionship, reducing loneliness and improving psychological well-being. Participants reported forming meaningful bonds with the robot and experiencing a sense of purpose in caring for it. These findings suggest that even if artificial emotion is merely simulated, its effects can be socially beneficial when properly implemented, ranging from emotional support and therapeutic assistance to enhanced user experience. As some optimists have stated \cite{fuchs2024understanding, duffy2003anthropomorphism}, 
humans are capable of deriving emotional satisfaction from simulations as we do with fictional characters or pets and without being deceived about their ontological status.

Moving towards modelling `inner' AI emotion as core component in the learning and decision making, i.e., turning towards artificial emotion, supporters think emotion may be essential for the development of higher intelligence. Some scholars argue that genuine intelligence is inseparable from emotional involvement. For instance, Marvin Minsky, one of the godfathers of AI, asserted, ``the question is not whether intelligent machines can have any emotions, but whether machines can be intelligent without any emotion''\cite{minsky1986society}. This position highlights the central role of emotion in cognitive decision-making. In humans, rational decisions often rely on affective signals to provide value-based assessments. Accordingly, some researchers suggest that without emotion-like mechanisms, AI may be unable to make complex and flexible decisions in a human-like manner. This argument has gained further traction in recent years:
Yann LeCun argues that machine emotions correspond to intrinsic cost signals and outcome anticipation by a trainable critic; hence autonomous agents will inevitably exhibit emotion-equivalents as part of their design \cite{lecun2022path}.  In this account, emotions functionally define what is preferable (goals) and how actions are evaluated (outcomes) via the cost module and critic, rather than being mere displays.
Similarly, neuroscientist Damasio et al.\ have advocated for embedding artificial equivalents of affect (e.g., homeostatic regulation mechanisms) in machines to support intrinsic motivation and self-preserving behaviour \cite{man2019homeostasis}. They propose that machines capable of sensing both their own internal states and others’ emotions (e.g., possessing empathic capacity) may not inevitably pose a threat to humanity. 

\subsection{Our Opinion on the Necessity of Artificial Emotion}
Following the technical and ethical reviews, we discuss 
our take on 
the necessity of artificial emotion. We propose that whether artificial intelligence should `have' emotion depends on its intended goals and application domains. For narrow or purely computational tasks, emotional `components' are not required; in fact, their inclusion may reduce objectivity and efficiency. However, if the objective is to build highly autonomous AGI or achieve more natural and intuitive human–AI interaction, some form of emotional mechanism may become necessary. Here, emotion refers not to unregulated human-like affective fluctuations, but to a controlled internal drive and value-evaluation system. Such a system can enable AI to respond to both environmental and internal states, maintaining goal-directed behaviour and adaptability across complex and changing contexts. 

Even so, we do not advocate for AI to 
necessarily  
replicate the full range of human emotional experience. From our perspective, the design and deployment of artificial emotion must be subject to strict safety and ethical constraints. First, the strength and type of emotional states should be limited. AI may exhibit mild affective states for decision modulation, such as suppressing harmful behaviour (akin to ``empathy''), or adjusting its strategy after repeated failure (akin to `frustration'). But it should not be driven by unregulated emotions such as unregulated `anger' or `fear' 
in the sense of rage or cold panic potentially leading to instable and uncontrollable states and outcomes.
Furthermore, 
we advocate a bounded emotional architecture design: preserve emergent affect when it remains within explicit safety constraints. While if an emotional response no longer aligns with predefined safety objectives, the system 
or external control mechanisms 
must be able to suspend or disable the relevant module. Such a bounded emotional architecture can allow AI to retain rational consistency while gaining the flexibility and adaptability typically associated with emotional reasoning.

Secondly, we strongly support the principle of transparency: AI's emotional mechanisms and expressions must be interpretable and explainable \cite{diaz2023connecting}
and users must be made aware of these. 
Both users and developers should be able to understand the reasoning behind an AI's emotional response, and be clearly informed 
Only through explicit communication can we prevent undue user trust. 
Indeed, many global AI guidelines stress the importance of transparency in affective systems, calling for clear disclosure of their limitations \cite{voigt2017eu}. Therefore, we recommend that AI interfaces include markers 
to continuously remind users of the artificial nature of the emotion, 
thus reducing the risk of relational illusions. At the development stage, explainability should also be embedded within the emotional algorithms, enabling external audits to verify their decision logic and ensure compliance with human values and interests, especially in sensitive sectors such as healthcare \cite{sun2025explainable}, finance \cite{george2023securing}, and biometric systems \cite{sun2024audio}.

\begin{table*}[t]
\centering
\small 
\caption{Technical functions of artificial emotion by architectural level, with status labels assigned by an self-defined explicit rubric.
\textit{Status legend:} \textbf{Absent} blank implementation; 
\textbf{Early} single-lab prototype or formative study without independent replication or cross-domain testing; 
\textbf{Partial} domain-specific evidence with at least one of: (i) independent replication, (ii) quantitative gains on public benchmarks, or (iii) deployment beyond toy settings; 
\textbf{Mature} widespread adoption (e.g., products or open-source toolchains) with third-party evaluations across $\geq$2 domains. 
Status labels reflect a dual coding by two authors (disagreements resolved by discussion).}
\begin{tabularx}{\textwidth}{
  >{\raggedright\arraybackslash}p{4cm}
  >{\raggedright\arraybackslash}p{6cm}
  >{\raggedright\arraybackslash}p{3cm}
  >{\centering\arraybackslash}p{1cm}
}
\toprule
\textbf{Emotion function} & \textbf{Description} & \textbf{Representative evidence} & \textbf{Status} \\
\midrule

\multicolumn{4}{l}{\textbf{Communicative Aspects}} \\
Label-conditioned affective generation & Controls output style or sentiment through explicit emotion tags supplied at inference time & Emotion recognition 
tasks~\cite{li2025gatedxlstm} & Mature \\
\addlinespace[3pt]
Social affect expression & Generates affective signals to maintain user engagement and well-being & LOVOT robot~\cite{tan2024improving} & Partial \\
\addlinespace[3pt]
Introspective affect reporting & Enables AI to self-monitor and truthfully report its own affect to external auditors & — & Absent \\
\addlinespace[3pt]
\midrule
\multicolumn{4}{l}{\textbf{Architectural Aspects — Display Models}} \\
\addlinespace[3pt]
Standardised evaluation metrics & Agreed quantitative benchmarks that compare affect modules across tasks & — & Absent \\
\addlinespace[3pt]
\midrule
\multicolumn{4}{l}{\textbf{Architectural Aspects — Process Models (Low-Level)}} \\
\addlinespace[3pt]
Reward-grounded affect & Converts task stakes into valenced signals that bias action selection and exploration & iCub battery “anxiety” prototypes~\cite{canamero2019embodied} & Early \\
\addlinespace[3pt]
Human-preference alignment & Uses crowd-rated affective quality as a reinforcement signal to steer generation & RLHF for LLMs~\cite{bai2022training} & Partial \\
\addlinespace[3pt]
Self-discovered affective latents & Learns internal affect dimensions via self-supervision; conditions downstream generation & RCG~\cite{li2024return} & Early \\
\addlinespace[3pt]
Salience-weighted memory & Writes high-valence events to slow-decay store and prioritises them during retrieval & Affective GWR~\cite{barros2017self} & Partial \\
\addlinespace[3pt]
\midrule
\multicolumn{4}{l}{\textbf{Architectural Aspects — Process Models (High-Level)}} \\
\addlinespace[3pt]
Control-layer modulation & Maps appraisal variables to symbolic weights for attention or strategy switching & NEMO selector node~\cite{costantini2024nemo} & Early \\
\addlinespace[3pt]
Homeostatic drives / intrinsic motivation & Synthetic interoception creates internal need states for self-preserving behaviour & — & Absent \\
\addlinespace[3pt]
Bounded-emotion safety governor & Supervisory circuit suppresses or disables modules if responses exceed safety bounds & — & Absent \\
\addlinespace[3pt]
\midrule
\multicolumn{4}{l}{\textbf{System-Level Integration}} \\
\addlinespace[3pt]
End-to-end loop in open-domain AGI & Fully integrates appraisal, memory, planning, expression, and learning & — & Absent \\
\addlinespace[3pt]
\bottomrule
\end{tabularx}
\label{tab:necessity}
\end{table*}

Furthermore, in considering the ethical and social implications of AE, we argue for a balanced approach that neither rejects the technology outright nor underestimates its risks. On the one hand, dismissing AE entirely may mean missing valuable opportunities. If deployed safely, 
AE 
can offer tangible benefits
including highly effective AGI that also could truly understand human values. 
One must carefully balance not missing on genuine benefits for mankind, but clearly, the concerns raised by critics must be addressed with serious attention at all times. As AE  continues to develop, it is essential to implement robust ethical and regulatory frameworks. 
Additionally, the emerging debate \cite{butlin2025principles, metzinger2021artificial}
around whether AI might one day experience 
human-like (or functionally equivalent) forms of suffering demands proactive ethical inquiry, to avoid unintentionally creating sentient systems that are capable of distress.

\begin{figure}
    \centering
    \includegraphics[width=0.8\linewidth]{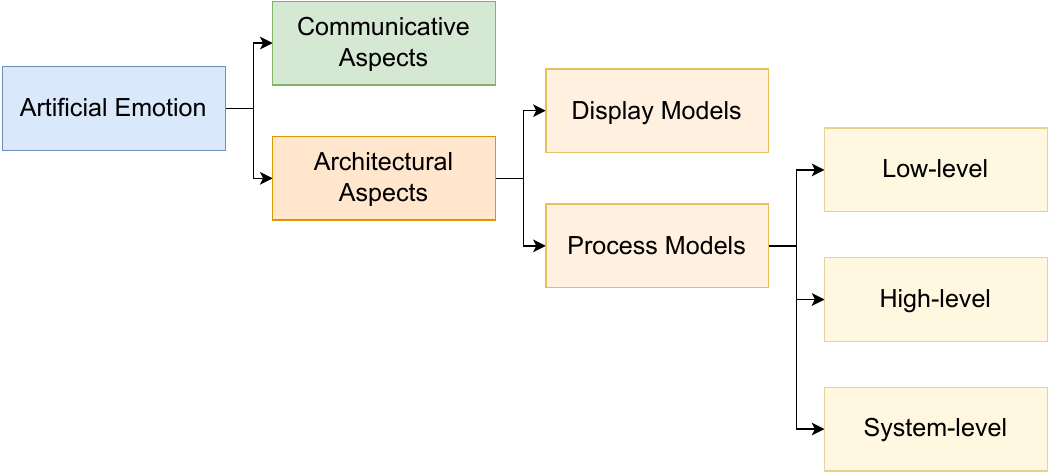}
    \caption{AE application architecture. The figure illustrates a hierarchical framework highlighting both existing applications and potential avenues yet to be explored.}
    \label{fig:AE}
\end{figure}

We remain optimistic about this line of research, as AE exhibits both established practical applications and unresolved research frontiers.
This is illustrated in Table \ref{tab:necessity}, which maps the discussed techniques onto the corresponding architectural levels, and further visualised in Figure~\ref{fig:AE}. Rather than treating it 
as a flat inventory, we can read it as a concrete recipe for how emotion is actually realised in machines. First, systems construct valenced signals that stand in for affect: reward-grounded schemes turn task stakes into anxiety‑ or relief‑like variables that bias exploration and action selection; preference‑based alignment (e.g., RLHF) supplies socially curated `felt value' instead of environmental grounding; and recent self‑supervised methods learn latent affect dimensions that subsequently condition planning and generation. Second, these signals are stored and prioritised: salience‑weighted memory architectures write high‑valence events to slow‑decay buffers and recall them preferentially, yielding the asymmetries we expect from affective cognition. Third, in emotion modulates control appraisal variables can be lifted to a symbolic control layer to re-rank goals, reweight attention, or trigger strategy switches; in a fully realised implementation, homeostatic drives would supply interoceptive needs and bounded‑emotion safety governors would cap intensity and shut modules down when limits are exceeded. Finally, affect is externalised and made auditable: label‑conditioned affective generation and socially oriented expression already operate at scale, while introspective affect reporting remains an essential but still-missing capability. 

Taken together, these steps show that artificial emotion is not `mystical' but implementable: 
for example, by 
a control-and-learning layer that derives valence, keeps a stateful, memory‑sensitive affect register, modulates perception, memory, planning, and communicates and constrains itself. Thus, whether AI should  have artificial emotion depends on our ambition. If we only need efficient tools, emotion is 
likely 
unnecessary. If we aim for autonomous, socially embedded AGI, a regulated, transparent, and safety‑bounded form of artificial emotion is likely indispensable. Only under these conditions can emotion become an asset rather than a liability in the development of AGI. Despite various interpretations and debates, it is important to approach the development of AE also with optimism rather than mostly fear. Given that research in this field is already advancing and existing techniques increasingly suggest the emergence of emotion-like processes in AI systems, halting progress out of concern risks to overlook its potential benefits and the inevitability of its continued evolution. 

%
\section{Conclusion}\label{sec6}
In this work, we have investigated AE within recent advancements in learning algorithms and system architecture, alongside a discussion of the ethical perspectives surrounding these developments. We gave a definition of AE and provided a blue print of how to realise artificial emotion in AI. We argue that AE is not only conceptually important but also vital for the progress of AGI. Moving forward, we advocate for greater attention to these domains, particularly in addressing critical challenges such as the absence of standardised evaluation metrics, the end-to-end loop in
open-domain AGI, the ethical implications,
and the need for transparency, interpretability, and accountability when applying AE. Although it is philosophically challenging to validate the existence of AI emotions---just as we cannot scientifically verify whether humans truly feel the emotions they express---we envision a future with 
emotion-aware AI research targeting friendly AI.

\bibliographystyle{plain}
\bibliography{reference}





\section*{Acknowledgments}
This research was partially supported and funded by the Munich Center for Machine Learning and the Munich Data Science Institute.

\vfill

\end{document}